\documentclass[aps,twocolumn,prl,superscriptaddress,amsmath,amssymb,amsfonts,nofootinbib]{revtex4-1} 

\usepackage[T1]{fontenc}
\usepackage[latin9]{inputenc}
\usepackage{lmodern} 
\usepackage{color}
\usepackage{bbold}
\usepackage{dsfont}
\usepackage{graphicx}
\usepackage[caption=false]{subfig}
\usepackage[colorlinks]{hyperref}
\usepackage{footnote}
\usepackage{enumerate}
\usepackage{float}
\usepackage{mathtools}

\usepackage{empheq}
\usepackage{tikz}
\usepackage{fancyhdr}
\usepackage[bottom]{footmisc}
\usepackage[normalem]{ulem}
\begin{document}

\global\long\def\eqn#1{\begin{align}#1\end{align}}
\global\long\def\ket#1{\left|#1\right\rangle }
\global\long\def\bra#1{\left\langle #1\right|}
\global\long\def\bkt#1{\left(#1\right)}
\global\long\def\sbkt#1{\left[#1\right]}
\global\long\def\cbkt#1{\left\{#1\right\}}
\global\long\def\abs#1{\left\vert#1\right\vert}
\global\long\def\der#1#2{\frac{{d}#1}{{d}#2}}
\global\long\def\pard#1#2{\frac{{\partial}#1}{{\partial}#2}}
\global\long\def\re{\mathrm{Re}}
\global\long\def\im{\mathrm{Im}}
\global\long\def\dd{\mathrm{d}}
\global\long\def\ddd{\mathcal{D}}

\global\long\def\avg#1{\left\langle #1 \right\rangle}
\global\long\def\mr#1{\mathrm{#1}}
\global\long\def\mb#1{{\mathbf #1}}
\global\long\def\mc#1{\mathcal{#1}}
\global\long\def\tr{\mathrm{Tr}}
\global\long\def\dbar#1{\Bar{\Bar{#1}}}

\global\long\def\nth{$n^{\mathrm{th}}$\,}
\global\long\def\mth{$m^{\mathrm{th}}$\,}
\global\long\def\non{\nonumber}

\newcommand{\orange}[1]{{\color{orange} {#1}}}
\newcommand{\cyan}[1]{{\color{cyan} {#1}}}
\newcommand{\blue}[1]{{\color{blue} {#1}}}
\newcommand{\yellow}[1]{{\color{yellow} {#1}}}
\newcommand{\green}[1]{{\color{green} {#1}}}
\newcommand{\red}[1]{{\color{red} {#1}}}
\global\long\def\todo#1{\yellow{{$\bigstar$ \orange{\bf\sc #1}}$\bigstar$} }
\global\long\def\pvcomm#1{\red{{$\bigstar$ \blue{\bf\sc [PV: #1]}}$\bigstar$} }
\title{Collective effects in Casimir-Polder forces}
\author{Kanupriya Sinha}
 \email{kanu@umd.edu}
\affiliation{Max Planck Institute for the Physics of Complex Systems, Dresden, Germany.
}
\author{B. Prasanna Venkatesh }
 \email{Prasanna.Venkatesh@uibk.ac.at}
\affiliation{Institute for Theoretical Physics, University of Innsbruck, A-6020 Innsbruck, Austria.}
\affiliation{Institute for Quantum Optics and Quantum Information of the
Austrian Academy of Sciences, A-6020 Innsbruck, Austria.}
\author{Pierre Meystre }
 \email{pierre@optics.arizona.edu}
\affiliation{Department of Physics and College of Optical Sciences,
University of Arizona, Tucson, AZ 85721, USA.}

\begin{abstract} We study cooperative phenomena in the fluctuation-induced forces between a surface and a system of neutral two-level quantum emitters prepared in a coherent collective state, showing that the total Casimir-Polder force on the emitters can be modified via their mutual correlations. Particularly, we find that a collection of emitters prepared in a super- or subradiant state  experiences an enhanced or suppressed collective vacuum-induced force, respectively. The collective nature of dispersion forces can be understood as resulting from the interference between the different processes contributing to the surface-modified resonant dipole-dipole interaction. Such cooperative fluctuation forces  depend singularly on the surface response at the resonance frequency of the emitters, thus being easily maneuverable. Our results demonstrate the potential of collective phenomena as a new tool to selectively tailor vacuum forces.
\end{abstract}

\maketitle

\textit{Introduction.}---Collections of atoms and solid-state quantum emitters coupled to waveguides and nanophotonic structures offer a promising platform for scalable quantum information processing  \cite{Nemoto2014, Lukin2001,  Yao2012, Hammerer2010}. The applications of such systems range from building long-ranged quantum networks \cite{Kimble2008, Nemoto2016}, quantum memory devices \cite{Asenjo2017, Sara2015, Steger2012}, and  metrology \cite{Zhou2014}, to facilitating new experimental regimes with exotic light-matter interactions \cite{Darrick2015, DEC2013}. When interfacing small quantum systems and surfaces at nanoscales, fluctuation-induced phenomena such as vacuum forces \cite{KSDEC}, surface-modified dissipation \cite{SpEm1, SpEm2} and decoherence \cite{spinflip3}, become an imperative element of consideration. The need to achieve better control and coherence of photonic systems at that scale requires therefore a detailed understanding of these phenomena, so as to determine the extent to which they can be tailored and controlled. In this work, we consider the possibility of using cooperative effects as a means to modify fluctuation-induced forces, or Casimir-Polder (CP) forces \cite{Milonni, CP1948}.

 \begin{figure}[b]
\begin{center}
\includegraphics[width = 3 in]{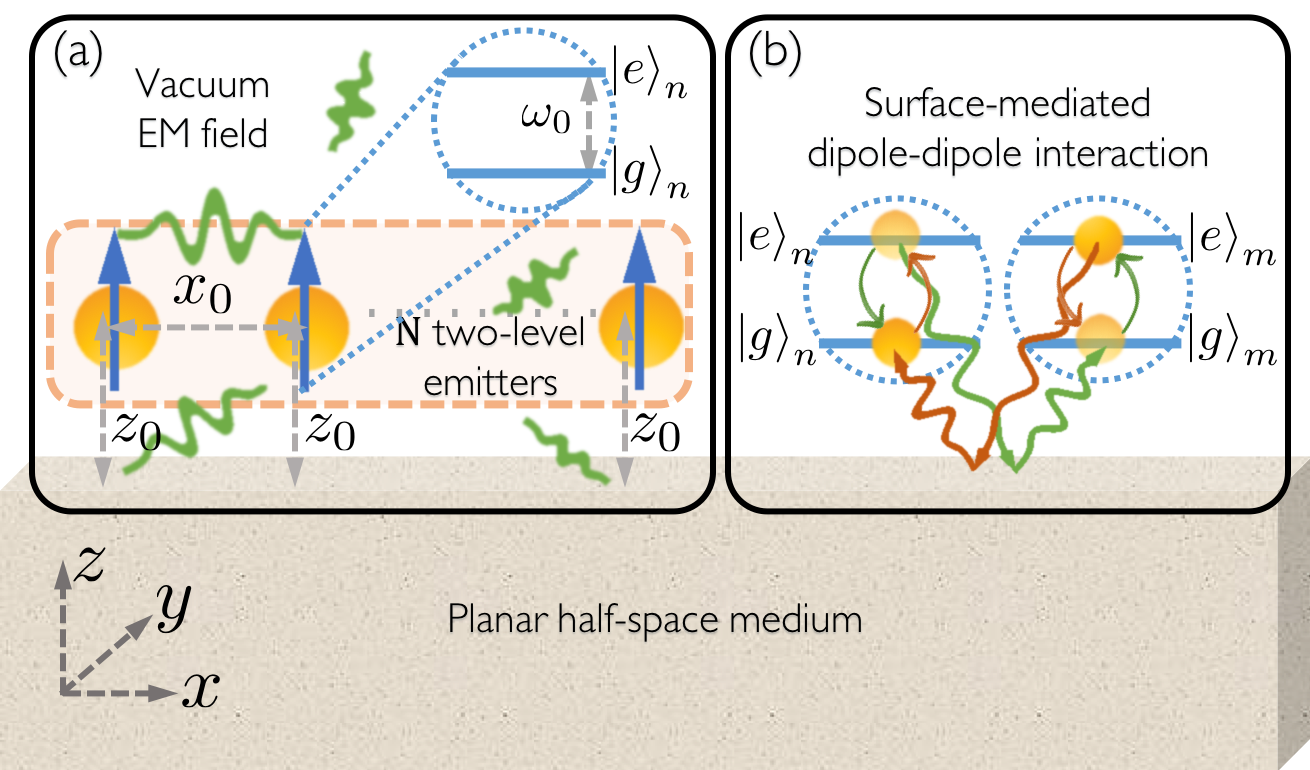} 
\end{center}
\caption{(a) Schematic representation of $N$ two-level quantum emitters prepared in a collective state, interacting with the vacuum EM field in the presence of a planar half-space medium.   (b) Constructive (destructive) interference between  the two processes shown in green and red leads to superradiance (subradiance) in the surface-mediated resonant dipole-dipole interactions.}
\label{schematic}
\end{figure}

The study of cooperative effects has a long theoretical and experimental history in the context of spontaneous emission from a collection of atoms in optical cavities and free space \cite{Dicke, Haroche, FicekTanas, MandelWolfBook, Andreev,  Feld1, DeVoe}, and more recently near waveguides \cite{Asenjo2017, Solano2017, Li2016}. Considering that  surface-modified spontaneous emission is the dissipative counterpart to the dispersive vacuum forces \cite{IntravaiaChapter}, one can expect to observe collective effects in dispersion forces as well.  When considering vacuum forces, however, the role of quantum coherence within or between the interacting bodies is seldom discussed. While there have been some investigations of the effect of correlations on the van der Waals forces between two atoms in a cavity \cite{Buhmann2017} and of interference effects in vacuum forces in a three level system \cite{collCP3}, a general analysis of fluctuation-induced forces between an $N$-particle system prepared in a coherent collective state and a macroscopic body is yet to be explored in detail. The goal of this letter is to analyze a proof of concept that illustrates cooperative effects in Casimir-Polder forces between a surface and a system of $N$ two-level quantum emitters prepared in a Dicke state \cite{Dicke}.

\textit{Model.}---We consider a one-dimensional chain of $N$ two-level quantum emitters at a distance $z_0$ from the surface of a  planar half-space medium, with each emitter separated by a distance $x_0$ from its nearest neighbor (see Fig.\,\ref{schematic}\,(a)). We assume that the half-space $z<0$ is occupied by a medium of dielectric permittivity  $\epsilon(\omega)$, while the upper half-space is vacuum. The  ground and excited levels for the $n^{\mr{th}}$ emitter are denoted by $\ket{g}_n$ and $\ket{e}_n$ respectively. The two levels are connected via an electric-dipole transition with resonance transition frequency $\omega_0$ and  spontaneous emission rate $\Gamma_0$, with $\hat\sigma_n^{+} = (\hat\sigma_n^{-})^\dagger = \ket{e}_{n} \bra{g}_{n}$ being the ladder operators for the corresponding transition. Defining the collective spin operators $\hat J_k  \equiv \sum_{n = 1}^N \hat{\sigma}_n^k$, the Dicke states $\ket{J, M}$, correspond to \cite{Dicke} \eqn{\label{dicke}\hat{\mb {J}}^2\ket{J,M} &= J(J+1) \ket{J,M},\text{\,and}\non\\
\hat{ {J}}_z\ket{J,M} &= M\ket{J,M}.}

The total Hamiltonian for the system of emitters and the electromagnetic (EM) field is $\hat H =\hat H_{S} + \hat H_F + \hat H_{\mr{int}}$, where 
${\hat H_S = \sum_ {n = 1}^N \hbar \omega_0 \hat \sigma_n^{+}\hat \sigma_n^{-}}$
is the  Hamiltonian for the two-level emitters, and $\hat H_F$ is the Hamiltonian for the medium-assisted EM field, which we assume to be in the vacuum state. 
 The electric dipole interaction Hamiltonian between the emitters and the EM field is 
${\hat H_{\mr{int}} = -\sum_{n = 1}^N \hat{\mb{d}}_{n}\cdot \hat{\mb{E}}\bkt{\mb{r}_n} }$, where ${\hat{\mb{d}}_n = \mb{d}_n \hat\sigma_n^{+}+\mb{d}^\ast_n \hat\sigma_n^{-} }$ is the electric-dipole operator for the $n^{\mr{th}}$ emitter and  $\hat{\mb{E}}\bkt{\mb{r}_n}$ is the  electric field at the position $\mb{r}_n$ of the $n^{\mr{th}}$ emitter in the presence of the surface (see \cite{Supplement} for further details). We assume the dipole moments of all the emitters $\mb{d}_n\equiv d_0 \mb{e}_z$ to be  equal in magnitude and aligned along the $z$-direction. 

The resulting dynamics of the density matrix $\hat \rho_S$ of the emitters, after tracing out the EM field, is described by the Born-Markov master equation \cite{BPbook}
\begin{equation}
\label{bmme}
\frac {d\hat\rho_S}{d t}  = -\frac{i}{\hbar} \sbkt{\hat H_S', \hat \rho_S} + \mathcal{L}'_S \sbkt{\hat \rho_S,}
\end{equation} where $\hat H'_S$ is the effective  Hamiltonian for the emitters in the interaction picture,
\eqn{\label{heff}\hat{H}_S' = &\hbar \sbkt{\sum_{n = 1}^N    {{\Omega_n^{(+)} }\hat\sigma_n^{+} \hat\sigma_n^{-}+\Omega_{n}^{(-)}  \hat\sigma_n^{-}\hat\sigma_n^{+} }+\sum_{m>n} \Omega_{mn}  \hat\sigma_m^{-}\hat\sigma_n^{+}}.} 
Here $\Omega_n^{(-)} =\frac{\mu_0 \omega_0}{\hbar\pi} \int_0^\infty \dd\xi\, \frac{\xi^2}{\xi^2+\omega_0^2}{\mb {d}_n^\ast\cdot {\dbar{G}_\mr{sc}\bkt{{\bf r}_n, {\bf r}_n,i\xi}}\cdot{\bf d}_n}$, and  $\Omega_n^{(+)} = -\Omega_n^{(-)}+\Omega_n^{(\mr{res})}$ are the Casimir-Polder shifts for the ground and excited states of the \nth  emitter, respectively. These shifts correspond to processes wherein the $n^{\mr{th}}$ dipole emits and reabsorbs a photon that is scattered off the surface, with the photon propagator given by the scattering Green's tensor  $\dbar{G}_{\mr{sc}} \bkt{{\bf r}, {\bf r}', \omega } $, which is defined as the solution to the homogeneous Helmholtz equation \cite{Buh1_Main, GreenWelsch_Main}
\eqn{\mb{\nabla}\times\mb{\nabla} \times\dbar{G}_{\mr{sc} }\bkt{\mb{r}, \mb{r}', \omega}&- \epsilon\bkt{\mb{r}, \omega}\frac{\omega^2}{c^2} \dbar{G}_{\mr{sc} }\bkt{\mb{r}, \mb{r}', \omega}=0.
}
Here $\epsilon\bkt{\mb{r}, \omega}$ is the space-dependent permittivity of the medium. Note that in addition to the broadband off-resonant contribution $\Omega_{n}^{(-)}$, the excited state has a resonant contribution \cite{Buh2_Main}
\eqn{\label{res} \Omega_n^{(\mr{res})}\equiv-\frac{ \mu_0 \omega_0^2}{\hbar} \re\sbkt{{\bf d}_n^\ast\cdot {\dbar{G}_\mr{sc}\bkt{{\bf r}_n, {\bf r}_n, \omega_0}}\cdot{\bf d}_n},}
that depends on the response of the environment at the transition frequency $\omega_0$ of the emitters.

 The surface-modified resonant dipole-dipole interaction frequency $\Omega_{mn}$ between the emitters $n$ and $m$ can be expressed as the sum of a contribution $\Omega_{mn}^{(\mr{free})}$ from the resonant exchange of excitation between the two dipoles via a photon propagating in free space, and a contribution $\Omega_{mn}^{(\mr{sc})}$ from a photon scattered off the surface, see Fig.\,\ref{schematic}\,(b), with~\cite{footnote1}
\eqn{\label{Ommn}\Omega_{mn}^{(\mr{sc}, \mr{free})} = -\frac{ \mu_0 \omega_0^2}{\hbar} \re\sbkt{{\bf d}_m^\ast\cdot {\dbar{G}_{\mr{sc},\mr{free} }\bkt{{\bf r}_m, {\bf r}_n, \omega_0}}\cdot{\bf d}_n}.}
Finally, the surface-modified Liouvillian is given by
\eqn{\mc{L}'_S\sbkt{\rho_S} =\sum_{m,n}\frac{\Gamma_{mn}}{2}\left(2\hat\sigma_m^{-}\rho_S\hat\sigma_n^{+}-\hat \sigma_m^+\hat\sigma_n^{-}\rho_S-\rho_S \hat \sigma_m^+\hat\sigma_n^{-} \right),}
where  $\Gamma_{nn}$ is the spontaneous emission rate for the excited state of the $n^{\mr{th}}$ emitter, and $\Gamma_{mn} = \Gamma_{mn}^{(\mr{free})}+\Gamma_{mn}^{(\mr{sc})}$ is the dissipative coupling coefficient between emitters $n$ and $m$, with
\eqn{\label{gamma}\Gamma_{mn}^{(\mr{sc}, \mr{free})}=\frac{2\mu_0\omega_0^2 }{\hbar} \, \im\sbkt{{{\bf d}_m^\ast\cdot \dbar{G}_{\mr{sc}, \mr{free}}\bkt{{\bf r}_m, {\bf r}_n, \omega_0}\cdot{\bf d}_n}}.
}
From Eqs.~\eqref{res} and \eqref{gamma} we have that the dissipative coefficients $\Gamma_{nn}^{(\mr{sc})}$ and $\Gamma_{mn}^{(\mr{sc,free})}$ are related to the resonant dispersive shift $\Omega_n^{\mr{(res)}}$, and the dipole-dipole interactions $\Omega_{mn}^{(\mr{sc,free})}$, respectively, by the Kramers-Kronig relations \cite{QVForcesBook}. As we show below, this implies that a collective enhancement/suppression of resonant van der Waals forces is concomitant with the cooperative behaviour of spontaneous emission.

\begin{figure}[t]
\begin{center}
\subfloat{\includegraphics[width = 0.9 \columnwidth]{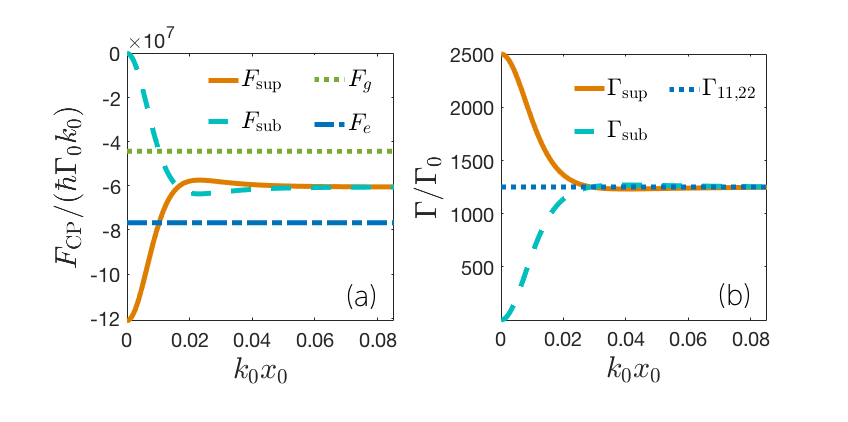}}\\
\vspace{-0.5 cm}
\subfloat{\includegraphics[width = 0.9 \columnwidth]{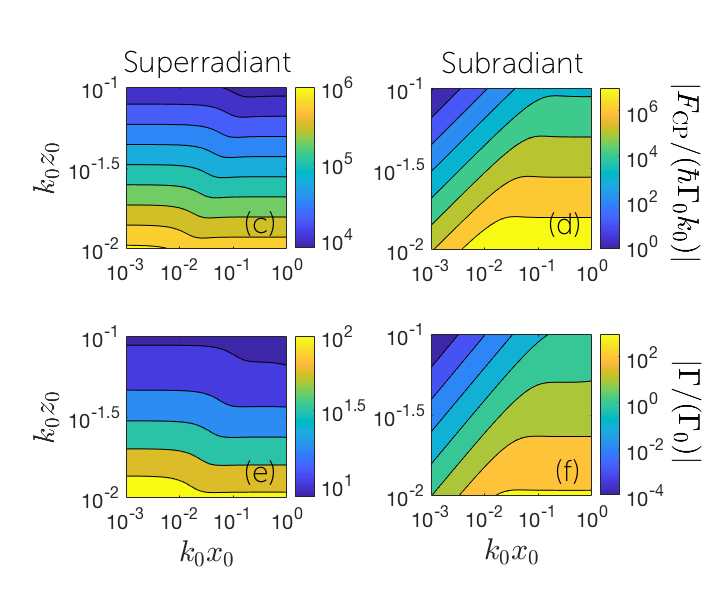}}
\end{center}
\vspace{-0.5 cm}
\caption{(a) Collective  Casimir-Polder force (in units of $\hbar \Gamma_0 k_0$), and (b) spontaneous emission, (in units of $\Gamma_0$),  on a system of two emitters near a gold surface, as a function of the separation between the emitters. Here the distance of the emitters from the surface is assumed to be $k_0 z_0 = 0.01$. (c) ((d)) Collective  Casimir-Polder force, and (e) ((f)) spontaneous emission on two emitters as a function of their distance from the surface and their mutual separation, for the dipoles prepared in  the superradiant (subradiant) state $\ket{\Psi_{\mr{sup}}}$ ($\ket{\Psi_{\mr{sub}}}$). The surface is described by the Drude model with a plasma frequency $\omega_p\approx1.36\times10^{16}$~Hz, and loss parameter $\gamma \approx 1.04\times10^{14}$~Hz for gold. }
\vspace{-0.5 cm}
\label{Fig2}
\end{figure}

\textit{Results.}---We define the total CP force for the system of emitters in a state $\hat\rho_S$ as $F_{\mr{CP}}\sbkt{\hat{\rho}_S} = -\pard{}{z}\tr\sbkt{\hat{H}_S' \hat{\rho}_S}$, so that
\eqn{
\label{fcp}
F_{\mr{CP}}\sbkt{\hat{\rho}_S} =&-\hbar \sum_{n = 1}^N\left [\frac{\partial}{\partial z }{\Omega_n^{(+)} }\avg{\hat\sigma_n^{+} \hat\sigma_n^{-}}+\frac{\partial}{\partial z }\Omega_{n}^{(-)}  \avg{\hat\sigma_n^{-}\hat\sigma_n^{+}} \right ] \nonumber \\
&- \hbar\sum_{m>n} \frac{\partial}{\partial z }\Omega_{mn}^{(\mr{sc})} \avg{\hat\sigma_m^{-}\hat\sigma_n^{+}+ \hat\sigma_n^{-}\hat\sigma_m^{+}},
}
 where all the averages are taken over the density operator $\hat\rho_S$. The first term corresponds to the CP forces on the individual emitters and the second term to the contribution from surface-modified dipole-dipole interactions. 
 
 Focusing on the second term in this expression we observe that while the operator average $\bkt{\avg{\hat\sigma_m^{-}\hat\sigma_n^{+}+ \hat\sigma_n^{-}\hat\sigma_m^{+}}}$ depends on the correlations between the dipoles in the state $\hat{\rho}_S$, the surface-modified dipole-dipole frequency $\Omega_{mn}^{(\mr{sc})}$ depends on the  average distance of the emitters from the surface. Hence, by preparing the emitters in a  suitable collective state $\hat{\rho}_S$, the CP force on an ensemble can be modified. Since this modification then depends only on the resonant frequency response of the surface, as evident from Eq.\,\eqref{Ommn}, it can thus be tailored easily by engineering surface resonances around the resonance  frequency of the emitters. This is the central message of this paper.

\begin{figure}[t] 
\begin{center}
\includegraphics[width = 0.9 \columnwidth]{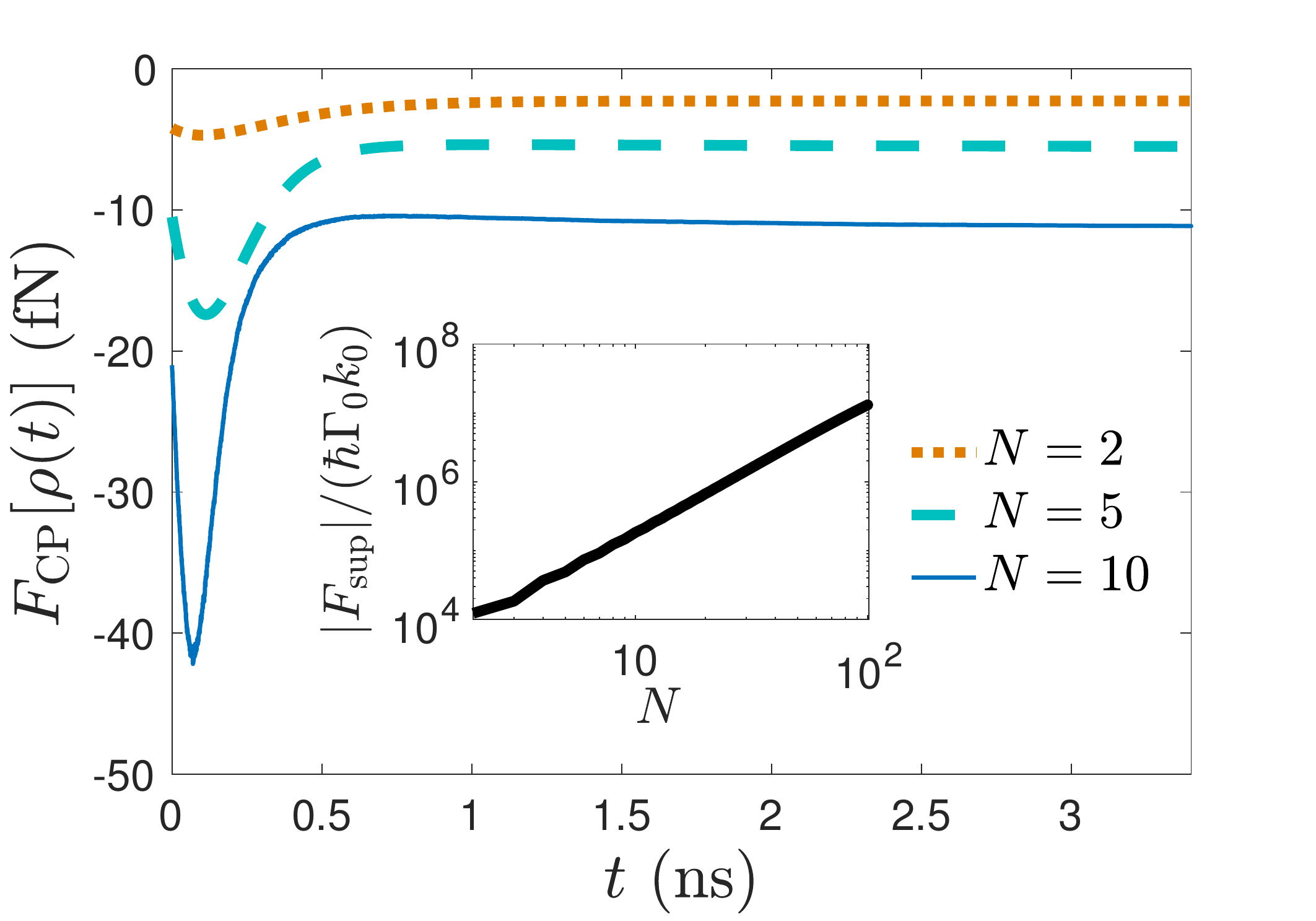}
\end{center}
\vspace{-0.5 cm}
\caption{Superradiant boost in the time dependence of the total attractive CP force for a collection of SiV emitters initially prepared in the excited level of the 737\,nm transition with a lifetime of 1.7\,ns \cite{Siv}. The emitters, with mutual separation $ x_0 \approx1 $\,nm, are assumed to be arranged in a linear chain inside a cantilever placed at a distance of $z_0 \approx 10$\,nm from a gold surface. The inset shows the absolute value of the maximum boost as a function of the number of emitters, illustrating the $N^2$ scaling of the force for the  superradiant state $\ket{J = N/2, M=0}$.}
\label{Fig3}
\end{figure}

As a first illustration consider two emitters prepared near a metal surface in one of the four internal states $\ket{\Psi_g}\equiv\ket{gg}$, $\ket{\Psi_e}\equiv\ket{ee}$, $\ket{\Psi_{\mr{sup}}}\equiv \bkt{\ket{eg}+\ket{ge}}/\sqrt{2}$, or $\ket{\Psi_\mr{sub}}\equiv \bkt{\ket{eg}-\ket{ge}}/\sqrt{2}$. We assume the surface to be described by the Drude model with permittivity $\epsilon\bkt{\omega} = 1 - \omega_p^2/\bkt{\omega^2 +i\omega \gamma}$, where $\omega_p$ and $\gamma$ are the plasma frequency and loss parameter for the metal, respectively. {From Eq.~\eqref{fcp} it follows that the force $F_{g(e)}$ for the state $\ket{\Psi_{g(e)}}$ is the sum of the forces on the individual emitters in the ground (excited) state, \begin{align}
    {F_{\substack{e\\g}} = -\hbar\sbkt{ \frac{\partial }{\partial z}\Omega_1^{(\pm)}+ \frac{\partial }{\partial z}\Omega_2^{(\pm)}} \approx-\frac{9\omega_p\hbar\Gamma_0 k_0 }{32 (\omega_p\mp\sqrt{2}\omega_0 )\tilde z_0^4}.}
\end{align} Here the approximate second expression corresponds to the non-retarded, or near-field, limit of the CP force valid in the emitters-surface distance regime ${\tilde z_0\equiv k_0 z_0 \ll1}$, with $k_0 \equiv \omega_0 /c$ \cite{CP1948, Buh1_Main, Supplement}. 

In contrast, the force on the super- and subradiant states, \eqn{F_{\substack{\mr{sup}\\\mr{sub}}} &= -\frac{\hbar}{2}\frac{\partial}{\partial z}\sbkt{ \Omega^{(\mr{res})}_{1}+\Omega^{(\mr{res})}_{2}\pm2 \Omega^{(\mr{sc})}_{12}},} includes a contribution that depends on the surface-mediated dipole-dipole interaction in addition to the resonant CP shifts of the individual emitters. In the non-retarded limit, it can be written as \eqn{\label{Fcollective}
&F_{\substack{\mr{sup}\\\mr{sub}}}\approx F_{\infty}\sbkt{1\pm f\bkt{\tilde x_0,\tilde z_0}},}
where we have introduced the asymptotic force for infinitely separated emitters 
\eqn{\label{Finfty} F_\infty \equiv  -\frac{9\omega_p^2\hbar\Gamma_0 k_0}{16\bkt{ \omega_p^2- 2\omega_0^2}\tilde z_0^4} ,}
and \eqn{f(\tilde x_0 ,\tilde z_0 )\equiv  \frac{8\tilde z_0 ^4}{3} \int_0 ^\infty\dd \kappa \kappa e^{-2 \kappa \tilde z_0}(\kappa^2+1)  J_0 \bkt{ {\tilde x_0}\sqrt{\kappa^2 +1} }}
quantifies the cooperativity due to geometric configuration of the dipoles, with $\tilde x_0 \equiv k_0 x_0 $. For coincident dipoles and to lowest order in $\tilde z_0$,  $\lim _{ x_0 \rightarrow 0 }f\bkt{\tilde x_0,\tilde{z}_0 } \approx1$. 

As illustrated in Fig.\,\ref{Fig2}\,(a), at small emitter separations $(x_0\lesssim z_0)$ the cooperative contribution leads to an enhanced and suppressed CP force for the super- and subradiant state, respectively. For larger separations, $\lim _{ x_0 \rightarrow \infty }f\bkt{\tilde x_0,\tilde{z}_0 } \approx 0$ and the interference effect in the resonant dipole-dipole interaction is attenuated, such that the super- and subradiant states experience an incoherent average of the ground and excited state forces, \emph{i.e.}, $F_{{\mr{sup},
\mr{sub}}}\approx \bkt{F_g+ F_e }/2 = F_\infty$. This is generally true for a state $\ket{\Psi_{\theta ,\phi}}\equiv \cos\theta \ket{eg}+e^{i\phi}\sin\theta \ket{ge}$ with a single shared excitation between the emitters. We note that the  total force on the state $\ket{\Psi_{\theta ,\phi}}$  is given by ${F}_{\theta,\phi} = - \hbar\frac{\partial}{\partial z} \sbkt{\Omega_1^{\mr{(res)}} +\Omega_{12 }^{(\mr{sc)}}  \sin(2\theta) \cos\phi}$, which can vary between the super- and sub-radiant values in Eq.\,\eqref{Fcollective}, depending on the relative amplitudes $(\tan\theta)$ and phase  $(\cos\phi)$ between the states $\ket{eg}$ and $\ket{ge}$. The collective spontaneous emission for the superradiant (subradiant) state, given by $\Gamma_{\mr{sup}} = 1/2\sbkt{ \Gamma_{11} + \Gamma_{22} + 2 \Gamma_{12}}$ $\bkt{\Gamma_{\mr{sub}} = 1/2\sbkt{\Gamma_{11} + \Gamma_{22} - 2 \Gamma_{12}}}$ is depicted in Fig.\,\ref{Fig2}\,(b) \cite{Supplement}.

In Fig.\,\ref{Fig2}\,(c)--(f), we provide a more comprehensive picture of the collective CP forces and spontaneous emission as a function of the geometrical configuration of the dipoles. Assuming the emitter resonant wavelength to be $\lambda_0 \equiv 2\pi c/\omega_0 \sim 700$\,nm, we see from  Fig.\,\ref{Fig2}\,(d) and Fig.\,\ref{Fig2}\,(f)  that a subradiant state of two emitters separated by $x_0\sim$1\,nm, and at a distance $z_0\sim$10\,nm from a gold surface, experiences a total force that is suppressed by a factor of $F_{\mr{sub}}/F_g\sim10^{-2}$ relative to the ground state van der Waals force, with a spontaneous emission $\Gamma_{\mr{sub}}/\Gamma_0\sim 10^{-2}$. Thus one can see that  subradiant CP forces can be a potential way to  avoid both dissipation and undesirable CP attraction.    

For a system of $N$ dipoles the CP force on the Dicke superradiant state $\ket{J=N/2,M=0}$ can be written as 
  \begin{align}
    F_{\mr{sup}} = -\frac{\hbar}{2}\sum_{n=1}^N \frac{\partial \Omega_n^{(\mathrm{res})}}{\partial z}- 2 \hbar \frac{\binom{N-2}{-1+N/2}}{\binom{N}{N/2}}\sum_{m>n} \frac{\partial \Omega_{mn}^{(\mr{sc})}}{\partial z} \label{eq:Fsp},
\end{align}
where $\binom{N}{k}$ is a binomial coefficient. In the limit of superposed dipoles, $x_0 \rightarrow 0$, it reduces to
\begin{align}
    \lim_{x_0 \rightarrow 0}F_{\mr{sup}} {=} -\frac{9\omega_p^2\hbar\Gamma_0 k_0}{32\bkt{ \omega_p^2- 2\omega_0^2   }\tilde z_0^4}\left(N+\frac{N^2}{2} \right),
\end{align}
which demonstrates the characteristic $N^2$ scaling of the collective CP force on the superradiant state, depicted in the inset of Fig.\,\ref{Fig3}, similar to free-space superradiant spontaneous emission at small emitter separations $\bkt{\tilde x_0 \ll 1}$ \cite{Haroche}. We also remark that, for $N>2$, multiple states in the degenerate subspace of subradiant Dicke states with $\ket{J=0,M=0}$ exhibit a suppressed CP force, see \cite{Supplement}.}

\textit{Discussion.}---We have identified  collective effects in vacuum-induced dispersion forces that  result from the interference between the different channels contributing to the surface-modified resonant dipole-dipole interaction, as sketched in Fig.\,\ref{schematic}\,(b). Such cooperative enhancement or suppression of fluctuation forces occurs for the resonant contribution to the total CP force, and can be physically understood as the dispersive counterpart to super- or subradiance in spontaneous emission (see Eq.\,\eqref{Fcollective}). In addition to the quantum correlations~\cite{footnote2} in the state of the emitters this contribution to the total CP force depends only on the surface response at the resonance frequency of the emitters, as can be seen from Eq.\,\eqref{Ommn}. It can be thus controlled by suitably tailoring the response of the surface around the resonant frequency of the emitters.  

Given that cooperative effects in optical dipole forces on solid-state emitters in nanodiamonds have been discussed both theoretically and experimentally \cite{Juan, Pras2018}, we suggest that it should be possible to observe a boost in the cooperative vacuum-induced forces by placing a similar nanodiamond doped with emitters near a  surface. To estimate the feasibility of observing the collectively enhanced CP force, we consider a system of $N$ Silicon-vacancy (SiV) centers embedded in a cantilever near a metal surface  \cite{Pellicione, Kleinlein}.  We assume that the emitters are initially prepared in the excited state, and solve the superradiance master equation, given by Eq.\,\eqref{bmme}, numerically \cite{Qutip13}. As the system decays in a collective manner, it occupies the superradiant manifold transiently and experiences an enhanced CP force, as shown in Fig.\,\ref{Fig3}. For a system of $N = 10$ SiV centers at a distance of $z_0\approx10$\,nm from a silica surface, we find a superradiant boost in the collective CP force of $\Delta F_{CP}\approx 20$\,fN over a time scale of $\Delta\tau\approx 0.5$\,ns. \cite{footnote3} While the magnitude of the enhanced force is large enough to be observable with current technologies \cite{NVsensor}, the time resolution required to sense the enhancement would appear to pose an experimental challenge.
 
 Alternatively, we note that Solano \textit{et al} \cite{Solano2017} have recently demonstrated cooperative effects in a collection of atoms placed near an optical fiber, wherein they exploited the van der Waals shifts to infer position of the atoms relative to the fiber. We remark that in such an experiment with atom-surface separations of $\approx30 $\,nm, the cooperative van der Waals shift for a  collection of $N = 6$ atoms in a superradiant state can be as large as $\sim 100$\,MHz, and can potentially provide an additional way of inferring the collective state of the atoms.

In terms of potential applications of collective vacuum forces, one can speculate that superradiant states could be used to boost and probe fluctuation forces that are otherwise too weak to be observable, as recently investigated in \cite{KS2017_main}.  Superradiant states of quantum emitters might also be a resource for sensing surface properties \cite{Ania2017} and quantum metrological applications \cite{Wang}. More interestingly perhaps, given that subradiant states suppress undesirable Casimir-Polder attraction and exhibit long lifetimes, they can be a useful resource for trapping particles near surfaces. 

\textit{Acknowledgements.}--- We are grateful to Ana Asenjo-Garcia, Ania C. Bleszynski Jayich, Mathieu L. Juan,  Francesco Piazza, Helmut Ritsch, Oriol Romero-Isart, and Pablo Solano  for insightful discussions. BPV is supported by the Austrian Federal Ministry
of Science, Research and Economy (BMWFW).

\clearpage
\onecolumngrid
\begin{center}

\newcommand{\beginsupplement}{%
        \setcounter{table}{0}
        \renewcommand{\thetable}{S\arabic{table}}%
        \setcounter{figure}{0}
        \renewcommand{\thefigure}{S\arabic{figure}}%
     }

\textbf{\large Supplemental Material}
\end{center}

\newcommand{\beginsupplement}{%
        \setcounter{table}{0}
        \renewcommand{\thetable}{S\arabic{table}}%
        \setcounter{figure}{0}
        \renewcommand{\thefigure}{S\arabic{figure}}%
     }

\setcounter{figure}{0}
\setcounter{table}{0}
\setcounter{page}{1}
\makeatletter
\renewcommand{\theequation}{S\arabic{equation}}
\renewcommand{\thefigure}{S\arabic{figure}}
\renewcommand{\bibnumfmt}[1]{[S#1]}
\renewcommand{\citenumfont}[1]{S#1}

\newcommand{\D}{\Delta}
\newcommand{\tD}{\tilde{\Delta}}
\newcommand{\K}{K_{PP}}
\newcommand{\bn}{\bar{n}_P}
\newcommand{\G}{\Gamma}
\newcommand{\LH}{\underset{L}{H}}
\newcommand{\HL}{\underset{H}{L}}

\section{Medium-assisted electromagnetic field and Green's tensor}
\label{appendixa}
Using the macroscopic QED formalism \cite{Buh1, Buh2}, the Hamiltonian for the vacuum EM field in the presence of the surface can be written as
\eqn{\label{hv}
\hat H_F =\sum_{ \lambda = e,m}\int \dd^3 r \int \dd\omega\,\hbar\omega\, \hat{\mb{f}}^\dagger_\lambda\bkt{\mb{r}, \omega}\cdot\hat{\mb{f}}_\lambda\bkt{\mb{r}, \omega},
}
with $\hat{\mb{f}}^\dagger_\lambda\bkt{\mb{r}, \omega}$ and ${\hat{\mb{f}}_\lambda\bkt{\mb{r}, \omega}}$ as the bosonic creation and annihilation operators respectively that take into account the presence of the media. Physically these can be understood as the ladder operators corresponding to the noise polarization ($\lambda = e$) and magnetization  ($\lambda = m$) excitations in the medium-assisted EM field, at frequency $\omega$, created or annihilated at position $\mb{r}$. The medium-assisted bosonic operators obey the canonical commutation relations \eqn{\sbkt{ \hat{\mb{f}}_{\lambda}\bkt{\mb{r}, \omega}, \hat{\mb{f}}_{\lambda'}\bkt{\mb{r}', \omega'} } &= \sbkt{ \hat{\mb{f}}^{\dagger}_{\lambda}\bkt{\mb{r}, \omega}, \hat{\mb{f}}^\dagger_{\lambda'}\bkt{\mb{r}', \omega'} }=0,\\
\sbkt{ \hat{\mb{f}}_{\lambda}\bkt{\mb{r}, \omega}, \hat{\mb{f}}^\dagger_{\lambda'}\bkt{\mb{r}', \omega'} } &8= \delta_{\lambda\lambda'}\delta\bkt{\mb{r} - \mb{r}'}\delta\bkt{\omega - \omega'}.}

The electric field operator evaluated  at the position of the $n^{\mr{th}}$ emitter is given as \eqn{\label{Era} \hat{\mb{E}}\bkt{\mb{r}_n}=& \sum_{ \lambda= e,m}\int \dd^3 r\int\dd\omega\sbkt{\dbar {G}_\lambda \bkt{\mb{r}_n, \mb{r}, \omega}\cdot \hat{\mb{f}}_{\lambda}\bkt{\mb{r},  \omega} + \text{H.c.}},}
where the coefficients $\dbar{G}_\lambda\bkt{\mb{r}_1,\mb{r}_2,\omega}$  are defined as 
\eqn{\dbar{G}_e \bkt{\mb{r},\mb{r}',\omega}=& i\frac{\omega^2}{c^2} \sqrt{\frac{\hbar}{\pi\epsilon_0}\im[\epsilon \bkt{\mb{r}',\omega}]} \dbar{G}\bkt{\mb{r},\mb{r}',\omega}, \\
\dbar{G}_m \bkt{\mb{r},\mb{r}',\omega}=& i\frac{\omega^2}{c^2} \sqrt{\frac{\hbar}{\pi\epsilon_0}\frac{\im [\mu \bkt{\mb{r}', \omega}]}{\abs{\mu\bkt{\mb{r}',\omega}}^2}}\nabla\times \dbar{G}\bkt{\mb{r},\mb{r}',\omega}.}
Here $\epsilon(\mb{r},\omega)$ and $\mu(\mb{r},\omega)$ refer to the space-dependent  permittivity and permeability, and $\dbar{G}\bkt{\mb{r}_1,\mb{r}_2,\omega}$ as the Green's tensor for a point dipole near a planar semi-infinite surface \cite{Buh1, Buh2, GreenWelsch}. The Green's tensor is defined as the solution to the Helmholtz equation in the presence of the boundary conditions
\eqn{
\mb{\nabla}\times \mb{\nabla}\times \dbar{G} \bkt{\mb{r},\mb{r}', \omega } - \frac{\omega^2}{c^2}\epsilon\bkt{\mb{r}, \omega}\mu\bkt{\mb{r}, \omega} \dbar{G}\bkt{\mb{r},\mb{r}', \omega} = \delta \bkt{\mb{r}-\mb{r}'} \mathbb{I}.
}

The total Green's tensor can be expressed as  
\eqn{ \dbar G\bkt{\mb{r}_1,\mb{r}_2,\omega} = \dbar G_{\mr{free}}\bkt{\mb{r}_1,\mb{r}_2,\omega}+\dbar G_{\text{sc}}\bkt{\mb{r}_1,\mb{r}_2,\omega} ,}
where $G_{\mr{free}}\bkt{\mb{r}_1,\mb{r}_2,\omega}$  and $G_{\mr{sc}}\bkt{\mb{r}_1,\mb{r}_2,\omega}$ refer to the free space and scattering components of the total Green's tensor. For a point dipole located at the position $\mb{r}_1$ near an infinite planar half-space, one can  write the scattering  Green's tensor as \cite{Buh1}
\eqn{\label{greensc}
  \dbar{G}_{\mr{sc}}\bkt{\mb{r}_1,\mb{r}_2, i\xi} &= \frac{1}{8\pi} \int_0^\infty \dd k_\parallel \frac{k_\parallel}{ \kappa_\perp } e^{- \kappa_\perp Z}\left[\left(\begin{array}{ccc}
    J_0(k_\parallel x_{12}) +J_2(k_\parallel x_{12}) &0 &0 \\
    0 &J_0(k_\parallel x_{12}) -J_2(k_\parallel x_{12})  &0\\
    0& 0&0
\end{array}\right)r_s\right.\non\\
&\left.-\frac{c^2}{\xi^2} \bkt{\begin{array}{ccc}
    \kappa_\perp^2 \sbkt{J_0(k_\parallel x_{12}) -J_2(k_\parallel x_{12}) }&0 &2k_\parallel \kappa_\perp J_1 (k_\parallel x_{12}) \\
    0 &\kappa_\perp^2 \sbkt{J_0(k_\parallel x_{12}) +J_2(k_\parallel x_{12}) }  &0\\
    -2k_\parallel \kappa_\perp J_1 (k_\parallel x_{12})& 0&2 k_\parallel^2 J_0 (k_\parallel x_{12})
\end{array}}r_p \right],
}
with $\abs{{\mb{r}_1-\mb{r}_2} }= r$, ${({\mb{r}_1+\mb{r}_2}) \cdot \mb{e}_z}= Z$, and we have defined the relative coordinate vector between the points $\mb{r}_1$ and $\mb{r}_2$ as ${\frac{{\mb{r}_1-\mb{r}_2}}{\abs{\mb{r}_1-\mb{r}_2}}\equiv \bkt{\frac{x_{12}}{r},0,\frac{z_{12}}{r}}^\mr{T}.}$ Here $r_{s,p}$ are the Fresnel reflection coefficients for the $s$ and $p$ polarizations reflecting off the surface, and $\kappa_\perp^2 = \xi^2/c^2 +k_\parallel^2$. Assuming that the medium can be treated as homogeneous and isotropic, and can be well-described in terms of its bulk optical properties at the length scales of the emitter-surface separations, we can consider that all the information about the surface material is accounted for in the following Fresnel  reflection coefficients
\eqn{\label{fresnel}
r_p\bkt{\kappa_\perp, i\xi} &=  \frac{\epsilon\bkt{i\xi}\kappa_\perp-\sqrt{\bkt{\epsilon\bkt{\mr{i}\xi}\mu\bkt{i\xi}-1}\xi^2/c^2+\kappa_\perp^2}}{\epsilon\bkt{i\xi}\kappa_\perp+\sqrt{\bkt{\epsilon\bkt{\mr{i}\xi}\mu\bkt{i\xi}-1}\xi^2/c^2+\kappa_\perp^2}},\non\\
r_s \bkt{\kappa_\perp, i\xi}&= \frac{\mu\bkt{i\xi}\kappa_\perp-\sqrt{\bkt{\epsilon\bkt{\mr{i}\xi}\mu\bkt{i\xi}-1}\xi^2/c^2+\kappa_\perp^2}}{\mu\bkt{i\xi}\kappa_\perp+\sqrt{\bkt{\epsilon\bkt{\mr{i}\xi}\mu\bkt{i\xi}-1}\xi^2/c^2+\kappa_\perp^2}}.
}
The free space Green's tensor between the points $\mb{r}_1$ and $\mb{r}_2$ is given as 
\eqn{
\label{greenfree}
\dbar{G}_{\mr{free}} \bkt{\mb{r}_1,\mb{r}_2, i\xi}& = \frac{c^2 e^{-\xi r/c}}{4\pi \xi^2 r^3} \bkt{\begin{array}{ccc}
     g\bkt{\frac{\xi r}{c}} - h\bkt{\frac{\xi r}{c}}\frac{x_{12}^2}{r^2}& 0&- h\bkt{\frac{\xi r}{c}}\frac{x_{12}z_{12}}{r^2} \\
     0&g\bkt{\frac{\xi r}{c}}&0 \\
- h\bkt{\frac{\xi r}{c}}\frac{x_{12}z_{12}}{r^2}     &0 &g\bkt{\frac{\xi r}{c}}- h\bkt{\frac{\xi r}{c}}\frac{z_{12}^2}{r^2}
\end{array}} .}
where $g\bkt{\chi}\equiv1+\chi+\chi^2$, 
$h\bkt{\chi}\equiv3+3\chi+\chi^2$.
\section{Superradiance master equation in the presence of a surface }
To find the influence of the medium-assisted EM field on the system of the two-level emitters, we derive the surface-induced modifications to the  master equation describing the dynamics of the corresponding reduced density matrix $\hat \rho_S$ \cite{KS2017}. Assuming that the dipoles are weakly coupled to the EM field, and that the  EM field bath correlations decay much faster than the relaxation time scale for the emitters' internal dynamics, we use the Born and Markov approximations to write the equation of motion for $\hat \rho_S$ as  \cite{BPbook_sup, KS2017}
\eqn{
&\der{\hat \rho_S}{t} =-\frac{1}{\hbar^2} \tr_F \int_0^\infty \dd\tau\sbkt{\tilde H_{\mr{int}}\bkt{t}, \sbkt{\tilde H_{\mr{int}}\bkt{t-\tau}, \hat \rho_S (t)\otimes\hat \rho_F}},
}
where $\tilde{H}_{\mr{int}}(t)\equiv e^{-i\bkt{\hat{H}_S +\hat{H}_F}t}\hat{H}_{\mr{int}}e^{i\bkt{\hat{H}_S +\hat{H}_F}t}$ stands for the interaction Hamiltonian in the  interaction picture. The reduced density  matrix $\hat{\rho}_F = \ket{0}\bra{0}$ refers to that of the vacuum EM field.  Tracing out the field, we obtain the surface-modified second-order Born-Markov master equation for the system dynamics as given by Eq.\,(2) in the main text. We observe that for a collection of coincident dipoles  $(x_0\rightarrow 0)$, one has  $\Omega_{ij} = \Omega_{kl}$ and  $\Gamma_{ij} = \Gamma_{kl}$, for all $\cbkt{i,j,k,l}$. It follows that the overall symmetry of the master equation remains the same as for  free space superradiance master equation \cite{Haroche_sup}, with  the Dicke states $\ket{J = N/2, M =0}$ and $\ket{J = 0,M=0}$ corresponding to  the super- and subradiant states for the $N$ emitters.

\section{Collective spontaneous emission near metal surface}
For two z-polarized dipoles at a distance $z_0$ near a metal half-space described by the Drude model with $\epsilon\bkt{\omega} = 1 - \frac{\omega_p^2}{\omega^2 + i \gamma\omega}$, the surface-modified spontaneous emission for the individual dipoles is given as 

\eqn{\Gamma_{nn}^{( \mr{sc})}=\frac{3 \Gamma_0 }{2 }\im&\sbkt{i \int_0 ^{1} {\dd \tilde k_\perp }{  } e^{2i \tilde k_\perp \tilde z_0 }\bkt{ 1 -\tilde k_\perp^2 } r_p\bkt{-ik_0\tilde k_\perp ,\omega_0 }+ \int_0^\infty {\dd \tilde\kappa_\perp }{} \bkt{1+ \tilde\kappa_\perp^2} e^{-2\tilde \kappa_\perp \tilde z_0} r_p\bkt{\tilde \kappa_\perp ,\omega_0 }},}
with $\tilde{z}_0 = k_0 z_0$. In the non-retarded limit, we get the modification to the dissipation as
\eqn{\label{Gnn}\Gamma_{nn}^{\mr{(sc)}}\approx \frac{3 }{8\tilde{z}_0^3 }\im \sbkt{\frac{\epsilon\bkt{\omega_0 }-1}{\epsilon\bkt{\omega_0 }+1}} \Gamma_0\approx \frac{3 \omega_0 \gamma }{4\omega_p^2\tilde z_0 ^3}\Gamma_0,
}
 which can be understood as the dissipative interaction between the $n^{\mr{th}}$ dipole and its image of strength  $\sbkt{\frac{\epsilon\bkt{\omega_0 }-1}{\epsilon\bkt{\omega_0 }+1}} \mb{d}$. We have  assumed here that $\omega_p \gg \cbkt{\omega_0 , \gamma}$. It can be seen that the for the chosen parameter values  in the main text, one gets $\Gamma_{nn}^{\bkt{\mr {sc}}}\sim  10^3 \Gamma_0 $, as can be seen from the blue dotted line  in Fig.\,2\,(b). For the surface modified dipole-dipole interaction with the dipoles separated by distance $x_0$, we have
 \eqn{\Gamma_{mn}^{( \mr{sc})}=\frac{3 \Gamma_0 }{2 }\im&\sbkt{i \int_0 ^{1} {\dd \tilde k_\perp }{  } e^{2i \tilde k_\perp \tilde z_0 }\bkt{ 1 -\tilde k_\perp^2 } J_0\bkt{\tilde k_\parallel \tilde x_0}r_p\bkt{-ik_0\tilde k_\perp ,\omega_0 }\right.\non \\
&\left.+ \int_0^\infty {\dd \tilde\kappa_\perp }{} \bkt{1+ \tilde\kappa_\perp^2} e^{-2\tilde \kappa_\perp \tilde z_0} J_0\bkt{\tilde k_\parallel \tilde x_0}r_p\bkt{\tilde \kappa_\perp ,\omega_0 }},}
with $\tilde{x}_0 = k_0 x_0$. In the non-retarded limit, this yields \eqn{\Gamma_{mn}^{\mr{(sc)}}\approx \frac{3}{2\tilde z_0 ^3} \im\sbkt{\frac{\epsilon(\omega_0- 1 } {\epsilon(\omega_0+ 1}} g\bkt{\tilde x_0 ,\tilde z_0 }\Gamma_0  ,} where $g \bkt{\tilde x_0 , \tilde z_0 }\equiv\int _0 ^\infty {\dd \kappa } \bkt{1 + \kappa^2 } e^{-2 \kappa \tilde z_0} J_0 \bkt{\sqrt{1+\kappa^2} \tilde x_0}$. In the limit of coincident dipoles ( $\tilde x_0 \rightarrow0$), this is equal to the single emitter spontaneous emission as given by Eq.\,\eqref{Gnn}. Thus, as  $\tilde x_0 \rightarrow0$, the super- and subradiant collective spontaneous emission rates are given as $\Gamma_{\mr{sup}}\approx2 \Gamma_{11}^{\mr{(sc)}}$, and $\Gamma_{\mr{sub}} \approx 0$, as can be seen from Fig.\,2\,(b) in the main text.

\begin{figure}[t]
    \centering
    \includegraphics{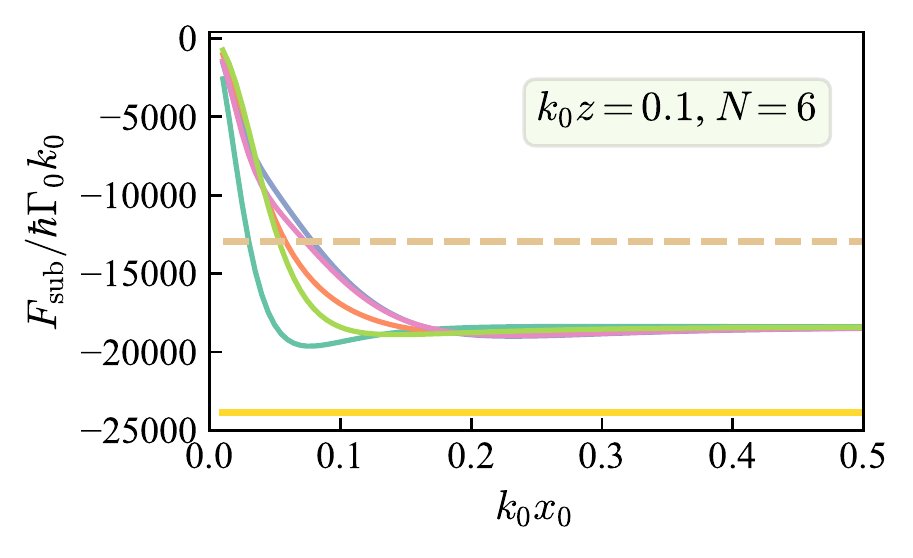}
    \caption{CP force on subradiant states of $N=6$ emitters placed at a distance of $k_0 z = 0.1$ away from a Gold surface as a function of their mutual spacing $x_0$. The horizontal lines give the force on the states with all the emitters in the excited (solid) and ground (dashed) states.}
    \label{fig:subR_supp}
\end{figure}

\section{$N$-emitter Subradiant states}
  
 Considering the subradiant state $\ket{J = 0,M=0}$, we first note that they have a degeneracy given by \cite{MandelWolf}
\begin{align*}
    d_G = \frac{N!}{\left(1+N/2 \right)!\,\left(N/2\right)!}.
\end{align*}
This degeneracy grows rapidly with $N$ and, in general, each subradiant state in this degenerate subspace has an intricate structure when expressed as a superposition in the product state basis over the energy eigenstates of the emitters. As a result, the general analytical expressions for the sub-radiant state CP force become cumbersome. Nonetheless,  for small numbers of emitters $N\sim10$, we have checked that all of the subradiant states demonstrate suppressed CP forces at small emitter separations and show the result calculated numerically for $N=6$ emitters ($d_G = 5$) in Fig.\,\ref{fig:subR_supp}.



\begin{thebibliography}{52}
\bibitem{footnote1} We have ignored the fourth-order van der Waals interactions between the dipoles, which scale as $\Omega_{mn}^{(4)}\sim1/x_0^6$, assuming that those contributions are weaker in comparison to the second-order interactions for the relevant dipole separations considered here $\bkt{k_0 x_0 \gtrsim (\Gamma_0 /\omega_0)^{1/3} }$.
\bibitem{footnote2} We remark that the resonant contribution to the total CP shift corresponds to the classical analog of a dipole interaction energy near a surface \cite{Hinds}, thus there should exist a similar cooperative dispersion force for a collection of classical dipoles with phase coherence radiating near a surface \cite{AntennaArray,Pustovit}.
\bibitem{footnote3} The numerical results for $N=10$ case are from a trajectory simulation averaged over $500$ trajectories, whereas for the smaller $N\leq6$, a direct simulation of the master equation Eq.~\eqref{bmme} was performed.
\end{thebibliography}
\end{document}